\documentstyle[12pt]{article}
\def\be{\begin{equation}}
\def\ee{\end{equation}}
\def\bea{\begin{eqnarray}}
\def\eea{\end{eqnarray}}

\begin{document}
\begin{center}
\Large
{\bf On Holography and Cosmology}
\end{center}
\vspace{.3in}
\normalsize
\begin{center}
{Reza Tavakol}$^1$  and 
{George Ellis}$^{1,2}$
\end{center}
\normalsize
\vspace{.6cm}
\begin{center}
{\em $^1$ Astronomy Unit, \\ School of Mathematical Sciences, \\
Queen Mary \& Westfield College,
\\Mile End Road, \\ London. E1 4NS. UK \\}
\vspace{0.3in}
{\em $^2$ Department of Applied Mathematics, \\ University of
Cape Town, \\ Cape Town, \\ South Africa}

\vspace{1cm}
SUBMITTED TO PHYS LETT B ON 22 JULY 1999

\end{center}
\vspace{1cm}
%________________________________________________________________

%________________________________________________________________
\begin{abstract}
%________________________________________________________________
Starting with the suggestion of Fischler and Susskind,
various attempts have recently been made to apply
the holographic principle to cosmology. Among these is a generalisation 
by Bousso which avoids the difficulties of the original proposal
in the re-collapsing FLRW models.

Here we take a closer look at the question of holography in 
cosmology with particular reference to this generalisation.
We demonstrate that in general realistic inhomogeneous universes 
such a proposal would involve extremely complicated - possibly 
fractal - light sheets. Furthermore, in a real inhomogeneous 
universe with evolving degrees of lumpiness on a variety of scales,
such a light sheet becomes time dependent
and cannot be known a priori on the basis of theory.
Its construction requires a detailed
knowledge of the distribution and growth of inhomogeneities
on all scales, and of the resulting caustics in 
null surfaces. Moreover, the evolution of the universe 
makes it clear that in general such bounds cannot remain
invariant under time reversal and will change with epoch.

We propose a modified version of this proposal in which
the light sheets end on the boundary of the past, and hence
avoid contact with the caustics. In this way the resulting 
light sheets and projections can be made much simpler. 
We discuss the question of operational definability of
these sheets within the context of both proposals and
conclude that in both cases the theoretical existence of
such sheets must be clearly distinguished from 
their complexity and the difficulty of their construction
in practice. This puts into perspective the likely 
practical difficulties one would face in applying the holographic 
principle to the real cosmos. These issues may also
be of relevance in debates regarding the applications of the 
holographic principle to other settings such as string theory.
%________________________________________________________________
\end{abstract}
%________________________________________________________________

%________________________________________________________________
\section{Introduction}
%________________________________________________________________
According to the so called {\it holography principle}, the maximum 
number of degrees of freedom in a volume is proportional
to its bounding surface  area \cite{thooft,susskind}. If true,
this would amount to an enormous simplification
of the world, as it would enormously reduce the
degrees of freedom required to understand it.
Furthermore, it would be informative as 
it could provide a {\it holographic bound} on entropy 
in a variety of physical settings, including cosmology.

There are two ways to approach this question:
either phenomenologically or at a fundamental level.
Here - in line with other applications of this principle 
to cosmology - we shall concentrate on the former.
We recall that an important motivation for this idea comes from
the Bekenstein-Hawking results \cite{bekenstein73,hawking} 
concerning black holes, according to which the entropy 
of the matter inside a black hole of mass $S_M$
cannot exceed the Bekenstein-Hawking entropy $S_{BH}$ given
by a quarter of the area $A$ of its event horizon in
Planck units, i.e.

\be
S_M \le S_{BH} = \frac{A}{4}.
\ee
The aim of the holographic principle is essentially to
generalise this result to more general settings, including
cosmology \cite{Fischler-Susskind}.  Leaving aside the justification for this
enormous extrapolation, this generalisation poses 
important questions. To begin with, as opposed to the
case of the black holes (BH), where appropriate notions of {\it volume}
and {\it surface} are naturally provided 
by the event horizon, it is not clear whether appropriate
analogues of these notions in fact exist in general cosmological settings
and if so whether they are unique and how they should be determined.
This is particularly true of the choice of surfaces, as even for a 
fixed volume, the surface is not uniquely defined. \\

Recently Fischler and Susskind \cite{Fischler-Susskind} 
have considered the application of holography principle to 
cosmology. Their proposal may be stated as follows:
\\

\noindent {\bf Fischler--Susskind Proposal} \cite{Fischler-Susskind}:
{\it Let $M$ be a four-dimensional spacetime.
Let $\Gamma$ be a spatial region in $M$ with 
a two-dimensional spatial boundary $B$.
Let $L$ be the light surface bounded by
$B$ and generated by the past light rays from $B$ towards the
centre of $\Gamma$. Then the entropy passing through
$L$ never exceeds the area of the bounding surface $B$.}
\\

In particular, 
in the case of adiabatic evolution,
the total entropy of the matter within the particle
horizon\footnote{Here this
is taken to mean the creation light cone, as in \cite{penxx},
rather than the set of particles bounding causal connection,
as originally defined by Rindler \cite{rin76,tipclaell}.}
must be smaller than the area of the 
horizon. They have shown that this proposal is
compatible with flat and
open Friedman-Lemaitre-Robinson-Walker (FLRW) cosmologies,
but that it fails for the $k=+1$
recollapsing models. 
A number of attempts have subsequently been made 
to remedy this difficulty 
\cite{reyetal}-\cite{rama-etal}.
In particular, Bousso \cite{bousso1,bousso2} has recently put forward 
a generalisation of this proposal and has applied it to a number of examples including
the recollapsing $k=+1$ FLRW cosmological models\footnote{See \cite{lowe}
for a recent critique of this scenario.}.\\

Here we take a closer look at the question of holography in a generic 
realistic inhomogeneous cosmological setting. We consider 
the proposal by Bousso as well as putting forward a modified version, 
in each case discussing the nature of the resulting light surfaces and
the difficulties in their operational definability.
In section 2, we briefly look at the application of this principle to 
FLRW models.  Sections 3 and 4 contain a brief discussion of Bousso's
proposal in the inhomogeneous cosmological settings, and the nature of the 
resulting light surfaces in these settings respectively. In section 5 we 
put forward a modified version to this proposal and discuss the nature
of the resulting light surfaces. Finally section 6 contains our conclusions.

%__________________________________________________________________
\section{Holography and FLRW universes}
%__________________________________________________________________
To begin with, let us briefly recall how the proposal
by Fischler and Susskind \cite{Fischler-Susskind} runs into difficulty
in the case of $K=+1$ recollapsing FLRW universes. In this case
the metric is given by
\be
ds^2 = -dt^2 +a^2(t) \left (d \chi^2 + \sin^2 \chi d\Omega^2 \right),
\ee
where $a(t)$ is the scale factor, $\chi$ is the azimuthal angle and 
$d\Omega^2$ is the line element of the 2--sphere at constant $\chi$. 
Assuming a constant\footnote{An assumption
that will obviously not be correct in a general inhomogeneous universe.} 
comoving entropy density
 $\sigma$, the ratio $S/A$ can be readily given as 
\cite{kaloperetal}
\be
\label{ratio}
\frac{S}{A} = \sigma \left [\frac{2 \chi_H - \sin 2 \chi_H}{4a^2 (\chi_H) 
\sin^2 (\chi_H)} \right ],
\ee
where $\chi_H = \int_0^t \frac{d t'}{a (t')}$ is the comoving
horizon size.
This clearly shows that the bound can be violated in this case, 
on noting that the area (given by the denominator of (\ref{ratio})) 
becomes  zero at the epoch of maximum expansion ($a = a_{max}$) 
given by $\chi_H =\pi$.\\

In order to remedy this shortcoming, Bousso has put forward a 
generalisation of the Fischler--Susskind Proposal which 
considers all four light-like directions and 
selects some according to an additional criterion of non-positive
expansion of the null congruences generating the null surfaces orthogonal
to the starting surface $B$. 
%He conjectures that in general cosmologies
%these restricted null surfaces store the entire corresponding bulk
%information, with a density of no more than one bit per Planck area.
More precisely, the idea is as follows: \\

\noindent {\bf Bousso's Proposal} \cite{bousso2}:
{\it Let $M$ be a four-dimensional spacetime which satisfies
Einstein's equations with the dominant condition holding
for matter. Let $A$ be the connected area
of a two-dimensional spatial surface $B$ contained in $M$.
Let $L$ be the connected part of a hypersurface bounded by
$B$ and generated by one of the four null congruences
orthogonal to $B$ such that the expansion of this congruence,
measured in the direction away from $B$, is non-positive everywhere. 
Let $S$ be the total entropy contained in $L$. Then $S \le A/4$.}\\
 
A hypersurface $L$ with the above properties is 
then referred to as a {\it light sheet (surface)} ${\cal{LS}}$ 
for the surface $B$. The crucial points regarding this 
proposal are that first it selects which of the four null 
surfaces orthogonal to $B$ can be considered as light sheets, and 
second it determines what part of those null surfaces will
be included in the light sheet: namely, they start at $B$, and 
any caustics present in the surface must act as end points to 
the light sheet \cite{bousso2}, if it is extended that far.\\

Considering the case of FLRW universes, 
on choosing a surface within the apparent horizon as the 
surface $B$, this proposal prevents the violation of the 
holography bound in the contracting phase \cite{bousso2}.
We note that in this simple (homogeneous and isotropic)
case the light sheet, given optimally by the
apparent horizon in a flat radiation dominated universe, 
is indeed connected as well as being
differentiable. Note that while the definition used is time 
symmetric, the null surfaces in the expanding universe 
are not invariant under time reversal except at an instant 
of maximum expansion in a homogeneous universe (which does 
not correspond to the present day situation), and clearly 
almost never in an inhomogeneous universe. \\

In the following sections we study various aspects of this
proposal in inhomogeneous cosmological settings. This is an 
essential extension of the previous work if it is to be taken 
as referring to the real universe.

%____________________________________________________________
\section{Holography and Inhomogeneous universes}
%____________________________________________________________
Let us consider a general realistic inhomogeneous universe, 
which may possibly possess a recollapsing phase.
To begin with, recall that a fundamental feature of classical 
self-gravitating systems is that in general they are not
in equilibrium states. This instability gives rise to
spontaneous creation of structure (lumpiness) which in the 
physically realistic case increases with time. As an example
of this we briefly recall a recent study of this
question in the context of Lemaitre--Tolman and Szekeres
inhomogeneous cosmological models \cite{mena-tavakol}.
Employing as a measure of density contrast (structuration),
covariant density contrast indicators in the form
\be                             
\label{index-rewrite}
DC=   \int_\Sigma  \left | \frac{h^{ab}}{\rho}
\frac{\partial \rho}{\partial x^a}                      
\frac{\partial \rho}{\partial x^b}\right | dV,
\ee
it has been shown that in general such structuration varies
with time, as expected\footnote{Moreover, it has been shown that
indicators of this kind exist which
grow monotonically with time for both
ever-expanding and recollapsing models
of Lemaitre--Tolman and Szekeres types, simultaneously
(see \cite{mena-tavakol} for details).}.
Here $\rho$ is the density, $h_{ab} = g_{ab} + u_a u_b $ projects orthogonal 
to the unit
4-velocity $u^a$ and $\Sigma$ is a 3-surface.
This indicates that 
density contrast (lumpiness) is likely to change with time
in inhomogeneous cosmological models, which is bound to be
reflected in the behaviour of the corresponding
Ricci and Weyl tensors.

Now the geometry of an arbitrary null surface encodes detailed information
about the Ricci and the Weyl tensors encountered by that surface, as shown
by the usual optical scalar equation (see e.g. \cite{wald})
\be
\frac{d \theta}{d\lambda} = -\frac{1}{2} \theta^2 - \sigma_{ab}  \sigma^{ab}
+ \omega_{ab} \omega^{ab} - R_{ab} \kappa^a \kappa^b
\ee
together with
\bea
\kappa^c \nabla_C  \sigma_{ab} &=& -\theta \sigma_{ab} + C_{cbad} \kappa^c \kappa^d \\
\kappa^c \nabla_C \omega_{ab} &=& - \theta \omega_{ab},
\eea
where $\theta$, $\sigma_{ab}$ and $\omega_{ab}$
are respectively the expansion, shear and the twist of a
congruence of null geodesics with the tangent field $\kappa^a$.
The convergence of the null generators
changes according to the null component of the Ricci tensor sampled
by the tangent vector and the rate of change of distortion according
to the Weyl tensor component sampled by these generators. The shear
in turn alters the expansion which determines the local rate of change
of area along the generators. The total area entering the inequality
in the holography conjecture 
        is a summation of all the resulting infinitesimal area
        elements of the null surface, and so 
              %in a sense 
        is a coarse-grained summation of all this information
        (in which all the fine details are lost).

As the universe evolves and structures form, the gravitational focusing 
(and caustic\footnote{Clearly focusing does not always lead to caustics.}) 
properties in inhomogeneous cosmologies are time dependent, which in turn 
makes the structure of the light surfaces time dependent
in these models, i.e. if we look at light surfaces associated with
a spatial surface $B_2$ that lies to the future of a surface $B_1$ 
we expect a time dependence in the associated area and entropy.\\

To fix ideas, consider an inhomogeneous universe possessing $N$ lumps 
at a given time $t$. Let us denote by $C_i$ the caustics produced by 
the lump $i$ in a null surface orthogonal to $B$ that starts off
with non-expanding normals at $B$. Now given that, according to 
the above proposal, caustics must act as end points to the light 
surface $\cal{LS}$, a necessary condition $\cal{LS}$ needs to satisfy 
to ensure the holography bound according to the above proposal is 
that it should contain all such caustics and hence
\be
{\cal{LS}} \supset {\bigcup}_i C_i.
\ee
This immediately raises a number of fundamental issues concerning the 
nature of such light surfaces and their operational definability 
in practice.

%--------------------------------------------------------------
\section{Nature of the light sheets in inhomogeneous universes}
%-------------------------------------------------------------
Assuming that the bound defined by $\cal{LS}$ does indeed 
hold in general inhomogeneous cosmological settings,
a number of important questions
still need to be addressed. They include:

\subsection{Differentiability and connectedness}
Given that light surfaces end at caustics,
their structures are in general forced to be extremely complex
and non-differentible, with possibly disconnected or even fractal
boundaries, depending upon the nature of
the inhomogeneous lumpiness and the resultant caustics in the universe.

We recall that a given source (lens) can
in principle produce a hierarchy of caustics with a range
of intensities. Thus each star will generally cause 
strong gravitational lensing with associated caustics\footnote{The
corresponding multiple images will in general not be detectable because
they will lie too close to the apparent surface of the star.}. 
However if the star
is in the core of a galaxy, there will also be much larger scale 
multiple images and caustics associated with the gravitational field of 
the galaxy itself; and if that in turn is in the core of a rich cluster of 
galaxies, the cluster will produce strong lensing with associated arcs and 
articles at even larger angular scales. In this way, each such a star
would contribute to multiple levels of lensing and caustics. Furthermore,
strong gravitational lensing is a commonplace
phenomenon. Indeed in the real universe we expect
such a hierarchical structure, with at least $10^{22}$ caustics
in our past light cone because of lensing caused by all the stars our
past light cone intersects in all visible galaxies, with further
multiple layers of caustics caused by additional lensing associated 
with at least some galaxy cores and some rich clusters of galaxies, as just
indicated. 

At each level, caustics occur that are associated with parts of
the past light cone that lie as indentations inside the boundary of the 
past, and are associated with multiple images of distant objects
(see for example \cite{Ellis-etal,Ehlers-etal}).
When lensing occurs, the past light rays generating a past light cone
self-intersect first non-locally and then locally, as one follows them back 
from the apex of the light cone. A light ray near a lensing object 
is deflected inwards by the gravitational field of the lens as it moves near 
it. It swings back 
towards the optic axis (the null geodesic from the observer through the
centre of the lens) and self-intersects a similar family of geodesics coming
from the other side. At this point it moves from generating the outer part
of the past light cone (the boundary of the past) to an inner part, folded
inside and lying within the past of the apex point\footnote{As is implied
by Figure 3 of \cite{bousso1}.}. It continues till
local self-intersection occurs at a cusp; from there on it generates the
back part of the folded light cone, 
%(see diagram 1)
which also lies
inside the past of the apex point. This is a general structure that results
from the nature of the boundary of the past of a set of points in a generic
space-time \cite{pen74}.
In the case of non-spherical lenses, multiple caustics due to a single
lens can lie inside each other; these complex nullcone
geometries have been
investigated analytically in the case of elliptical lenses, and numerically
in the case of realistic lensing models (see e.g. \cite{ber98}).

Additionally, it has been shown that the presence of single BH can 
produce an infinity
  %\footnote{Of course in quantum mechanical setting this number will 
  %be bounded by the the number of distinct light rays one can consider.},
of caustics associated with light rays that circle the black hole
an arbitrary number of times - albeit with rapidly decreasing 
intensities\footnote{Note that even though {\it in practice} caustics 
below a certain cut-off intensity may be ignored, in principle they
all need to be taken into account.} \cite{ellis-etal,corley-jacobson}.
Thus even a finite universe with say $10^{11}$ black holes in the visible
region - a very conservative estimate, given that we expect massive
black holes at the centres of many galaxies, as  well as all those resulting
from the collapse of super-massive stars - could give rise to an infinite 
number of caustics associated with each of these black holes, and hence
to an extremely complicated light surface. 

In this way the light sheet $\cal{LS}$ (whether the past light
cone of single point or not) may be said to 
{\it light trace} the content of the universe on all scales and thus 
encode its complexity, particularly through its caustic structure. 
This makes sense, as in contrast to the case of BH\footnote{Where the 
presence of the no-hair theorem `smoothes' the information
on the event horizon boundary.} and completely smooth
FLRW cosmological models, for 
which $\cal{LS}$ is readily given in terms of the small number of 
parameters which characterise these systems\footnote{Namely
mass $M$ in the case of BH and the deceleration and the Hubble 
parameters ($q_0, H$) in the case of FLRW models.}, one would in 
general cosmological settings expect this surface to be complex 
since no such constraints exist. To describe the detailed structure 
of a null surface (e.g. a past light cone) in a realistic cosmological 
setting will require many millions of parameters.

\subsection{Operational definability}

Strictly speaking, to define the holography bound precisely, all $C_i$ 
need to be included in the construction of $\cal{LS}$. The problem, 
however, is that the details of $C_i$ are not given a priori in terms of
theory, but depend on the details of the contents of the universe
(including the masses and sizes of the sources and lenses,
together with the detailed knowledge of their distributions
in space and time), which needs to be specified through observations.
The crucial point being that $\cal{LS}$ is constructive rather than 
theoretically given. This then raises the important question of
operational definability of $\cal{LS}$ for the real universe. Now 
given that all observations possess finite resolutions,
only sources, lenses and caustics above certain threshold levels
can be observable in practice. In this way,
a cut-off (course-graining) is inevitably involved
in the definition of $\cal{LS}$. Thus limitations in observational 
resolution become a barrier to constructing
the precise form of $\cal{LS}$ and hence ensuring the bound.

This then raises the interesting question of
whether one could formulate an {\it averaged holography principle}
in terms of the averaged (coarse-grained) light surface $\cal{ALS}$.
The problem, however, is that the coarse-graining (averaging)
of the content (say matter distribution)
does not commute with the averaging of the geometry,
mainly due to the nonlinearity of the curvature 
tensor (see for example \cite{tavakol-zala} and
references therein for a detailed discussion of this issue).
Worse still, coarse-graining of neither of these two quantities
would in general commute with coarse-graining the caustics. 
For example, a single BH of a given mass can produce 
an infinite number of caustics; it is not clear that a cut-off on 
the mass of the lens in general results in a similar cut-off
in the area of the resulting caustics.

\subsection{Time dependence and reversibility}

As was pointed out above, in a real universe the number of lumps $N$
as well as their positions, masses and shapes vary with time,
This is interesting as, in addition to the precise distribution
of sources, lenses etc, their time evolution is also important
for the construction of $\cal{LS}$, which is time dependent as the
surface $B$ is moved to different epochs in the universe's history.
Now given that this evolution must ultimately be related to the question 
of entropy (whatever its precise formulation may be in the presence of 
gravitational fields), the surface $\cal{LS}$ seems to encode 
time-dependent information regarding entropy as well. What has not been
done is to show that at later times in a realistic cosmological setting the
total entropy encoded this way will be larger than at earlier times. 
This is one of the  issues that needs clarifying; if the definition of 
entropy has the  desired properties, this must work out successfully. 
 %This has not yet been shown in the FLRW case, much less in the more 
 %realistic case considered here.

Additionally, even though 
locally (in a spacetime sense) in the neighbourhood of the bounding
surface $B$ one might argue (as is done by Bousso \cite{bousso1})
that the screen {\it definition} is invariant under time reversal, 
the actual surfaces will not be in an expanding universe: a unique 
direction of time will be picked by the expansion of the universe, 
and usually this will be marked by a difference in the expansions 
of the null normals to $B$. This difference will be enhanced by
a major difference between the caustics encountered in the future 
and the past of $B$, with the growth of inhomogeneities at quite 
different evolutionary stages in the two directions of time from
$B$, giving another way in which the geometry
of $\cal{LS}$ causes this time symmetry to be violated. 

%_______________________________________________________________
\section{A modified proposal}
%_______________________________________________________________
In the previous section it was shown that the light surfaces in 
Bousso's proposal are likely to have extremely complicated
structures in a real inhomogeneous universe. This is a direct 
consequence of the fact that in this proposal light surfaces are 
taken to end at caustics. 

Here we put forward a modified proposal which drastically simplifies 
the structure of these light surfaces. 
Before doing so, we note that it is important to distinguish carefully 
between ${\cal{P}_B}$, the boundary of the past of the 2--surface $B$
that starts off from $B$ with converging null geodesics, 
and the light sheet $\cal{LS}$ suggested by Bousso.
The former is a subset of the latter; the boundary ${\cal{P}_B}$ ends 
at the first self-intersections of the null sheets orthogonal to $B$, 
which will usually be non-local intersections, whereas the light sheet 
ends at caustics, which are local self-intersections.  It is the region
	between these self-intersections where the complex connectivity occurs. 

It therefore makes sense to separate out the part of the light surface 
which is not part of the boundary of the past of $B$. We shall call this
the {\it inner light sheet} ($\cal{ILS}$) and refer to the 
rest of the $\cal{LS}$, i.e. the part of the null surface through B 
that is also the null boundary of the past of $B$, as the {\it outer 
light sheet} ($\cal{OLS}$).
Then $\cal{ILS}$ encodes detailed information on the strong
lensing that occurs for the null surface, for it bounds the region after
self-intersection but before caustics. The number and topology of such
components depends on the lensing objects and hence reflects the degree
of strong density inhomogeneity. However, weak inhomogeneities will not
cause strong lensing and so will not be encoded in $\cal{ILS}$.
It is thus this inner light sheet that produces the enormous complexity 
in the light sheets proposed by Bousso. 

Additionally, continuing the null surface beyond the first 
self-intersections until the caustics actually result in multiple coverings
of part of the interior of $B$ by the null surfaces, which have turned
in on themselves. Thus including this part results in excess area
being counted, and a much more complex projection of data onto the
null boundary than is necessary when setting up the holographic
principle. For this purpose, it is only necessary to include data on 
the outer light surface $\cal{OLS}$; the data on $\cal{ILS}$ is then
redundant, having already been counted on $\cal{OLS}$.

We therefore propose a modified version of Bousso's proposal thus:\\

\noindent {\bf Proposal:}
{\it Let $M$ be a four-dimensional spacetime which satisfies
Einstein's equations with the dominant condition holding for matter.
Let $A$ be the connected area of a two-dimensional spatial surface $B$ 
contained in $M$. Let $L$ be the hypersurface bounded by
$B$ and generated by one of the four null congruences
orthogonal to $B$ such that the expansion of this congruence,
measured in the direction away from $B$, is non-positive everywhere, 
and ending on the boundary of the past of $B$.
Let $S$ be the total entropy contained in $L$. Then $S \le A/4$.}\\

The hypersurface $L$ with the above properties is
the outer light surface $\cal{OLS}$. 
The important feature of this modified proposal is 
that it cuts out the $\cal{ILS}$, together with the caustics
and the fractal boundaries arising from
them, and therefore has a much simpler light sheet structure.
It also covers regions in the interior of $B$ only once. We therefore 
suggest that, in the case of realistic
inhomogeneous cosmologies, this is a better surface to choose 
for the holographic principle
and the associated entropy conjecture.

%--------------------------------------------------------------
\subsection{Nature of the light sheets in the modified proposal}
%-------------------------------------------------------------
To begin with let us note that the modified covariant entropy
conjecture proposed above leaves unchanged all
the examples considered by Bousso \cite{bousso2}, including 
the $K=+1$ FLRW model. This is clear since none of the null
surfaces in these examples contain self-intersections 
    other than caustics, for example those at the origin of coordinates 
   ($r = 0$) in the light-sheets of 2-surfaces $B$ that
    are spherically placed about the origin (see Figure 2 
    of \cite{bousso1}.)

In the case of the more complicated inhomogeneous models
with both caustics and self-intersections present, on the other hand, 
replacing $\cal{LS}$ by  ${\cal{OLS}}$ enormously simplifies the 
structure of the light surface. Despite this, there are
a number of difficult points that still remain.\\

Firstly, even though the caustics are removed in this
formulation, the non-continuity of the generators of the boundary of the 
past of $B$ still remain at the self-intersections, which make the 
${\cal{OLS}}$ surface non-differentiable there. However a simple smoothing
over these regions where the outer surface has self-intersections 
should deal with this adequately in most cases, 
where the area of the smoothed surface can be arbitrarily close to that of 
the real surface. When this smoothing cannot be done, new effects
may occur and a very careful analysis will be required.

Secondly, since caustics can be locally determined,
the end points of the ${\cal{LS}}$ in Bousso's proposal were definable 
locally - at least {\it in principle}. Our proposal
relies on the determination of the  null boundary of the past of $B$,
which cannot be determined locally. So in this sense
there are both advantages and disadvantages
with this new proposal. It greatly simplifies the 
shape of the ${\cal{LS}}$ - at least in theory, but operationally 
it is still difficult to determine ${\cal{OLS}}$.

\subsection{Coarse-graining and information loss}
Crucial to the whole discussion is the issue of the scale of 
description used in the space-time model. One can
represent the same physical situation at different averaging scales:
thus we can represent the real universe (a) at a smoothed out
cosmological scale, where a FLRW model will suffice; (b) a finer
scale, where each cluster of galaxies is represented as an inhomogeneity;
(c) a finer scale, where each galaxy is represented; (d) a still finer
scale where each star and each black hole is individually represented.
The nature of the surface ${\cal{LS}}$ will be drastically 
different in these different representations. A coarse-graining
procedure will relate them to each other \cite{ell84} - remembering all
the time that these are different geometrical representations of the 
same physical situation.

Now it is plausible that in most circumstances, the definition of
entropy is closely associated with coarse-graining, and with the loss of
information that results from coarse-graining \cite{pen89} (see
also \cite{brandenberger-etal}). We might 
therefore expect quite different results for the entropy determined 
in terms of the areas associated with the null surfaces obtained
on different averaging scales for models representing the same
physical situation. We regard this as a fundamental issue but will
not pursue it further here except for the following remarks.

The area of the smoothed out model can be expected to be close to 
that obtained in the detailed (lumpy) model for the surface ${\cal{OLS}}$.
It is the area of ${\cal{LS}}$ and of ${\cal{ILS}}$ that will be very
different in these two cases; indeed the latter will be empty in the
smoothed-out case. The area of ${\cal{ILS}}$ is associated with strong
lensing only, and may perhaps be considered as a measure of the
pure gravitational entropy of the solution: the larger that area is, the
larger the degree of inhomogeneity (and hence entropy)
encoded in the gravitational 
field. On coarse-graining and consequent smoothing of the matter
representation, the corresponding ${\cal{ILS}}$ will decrease to zero.
The loss of this area represents loss of detailed information on the
inhomogeneity structure of the gravitational field resulting from this
coarse-graining. The one-way nature of the information loss associated
with coarse-graining is reflected in the fact that the area of the 
fine-grained surface ${\cal{LS}}$ is necessarily larger than that of 
${\cal{LS}}$, the latter being close to the area of the ${\cal{OLS}}$ 
in the smoothed-out (coarse-grained) description. 

There are potential parallels here with 
the presence of reversibility at
the level of microphysics and irreversibility at the
macroscopic level. What is needed to make the definitions and theory 
compelling is a comparison of entropy estimates and associated areas
at earlier and later times in the history of the expansion of the
universe, at different scales of description. 
We do not attempt this here, but note it as an important problem.  

%_______________________________________________________
\section{Conclusion}
%_______________________________________________________
We have taken a closer look at the applications
of the holography principle to cosmology and in
particular the proposal recently put forward by Bousso.
We have argued that in a real inhomogeneous universe,
the light surfaces defined in his way 
in order to satisfy the holography principle
would be non-differentiable and 
extremely (in principle infinitely) complicated.
Such a light surface can be viewed as a {\it light tracing} of the
complexity in the universe, projected onto this surface; like a 
cosmological analogue of the
images on the walls of Plato's cave!

In this way, to satisfy the holography principle in a
general inhomogeneous universe requires a detailed 
knowledge of the contents of the universe
and in turn its detailed caustic structure. Furthermore, the inevitable 
limits to the observational resolution puts fundamental limits on the 
operational definability of this surface.
Moreover, given the dynamical (and irreversible) evolutionary nature of 
light surfaces in general, such bounds cannot remain
invariant under time reversal.
 
We have introduced an alternative proposal
which results in a much simpler light surface.
However, operationally, it is still very difficult to 
define such surfaces in practice.

This leads us to conclude 
that in a realistic setting the theoretical existence of
such surfaces must be clearly distinguished from
their complexity and operational definability.

It would be extremely useful if an {\it averaged holography
principle} could be formulated.
Failing this, given the enormous amount of detailed information 
required for the construction of such light sheets, 
it is difficult to see how such a principle 
- formulated phenomenologically - can prove useful 
in simplifying the understanding of the cosmos
in {\it practice}. 
This of course does not rule out the possibility 
of correctness and usefulness of such a principle 
in the real world at a fundamental level and that 
could still be vitally important.
On the other hand, the phenomenological difficulties 
raised here, including the complexity, non--differentiability and potential
fractality of such surfaces,  might 
have some relevance in debates regarding 
the applications of the holographic principle 
at a fundamental level in other settings such as string theory. 

\vspace{.1in}
\noindent {\bf Acknowledgments:}
We wish to thank Robert Brandenberger, Nemanja Kaloper and Lee Smolin for
valuable discussions and comments. 
RT benefited from PPARC UK Grant No. L39094 and GE from support from the
FRD (South Africa).
%________________________________________________________________

%_______________________________________________________________ 

%___________________________________________________________
%___________________________________________________________
\end{document}